\documentclass[final]{elsarticle}
\usepackage[T1]{fontenc}
\usepackage[utf8]{inputenc}
\usepackage{lmodern}

\usepackage{amsmath,amssymb,amsfonts}
\usepackage{bm}
\usepackage{mathtools}

\usepackage{graphicx}
\usepackage{float}
\usepackage{subcaption}

\usepackage{booktabs}
\usepackage{multirow}

\biboptions{numbers,sort&compress}

\usepackage[
colorlinks=true,
linkcolor=blue,
citecolor=blue,
urlcolor=blue
]{hyperref}

\usepackage{siunitx}

\usepackage{enumitem}
\usepackage{physics}

\usepackage[margin=2.5cm]{geometry}

\begin{document}
\begin{frontmatter}
\title{\textsc{\textbf{Dual-Layer Optical Security Framework for Cryptogram Camouflage Using Circular Harmonic Components}}}

\author [1]{Jorge-Enrique Rueda-P \corref{cor1}}
\ead{jruedap2003@unipamplona.edu.co}

\author[1,2]{Carlos Pinzón}
\cortext[cor1]{Corresponding author} 

\affiliation[1]{organization={ Grupo de investigación Óptica Moderna (GOM)-MINCIENCIAS-COL, Departamento de Física, Facultad de Ciencias Básicas},
addressline={Universidad de Pamplona},
postcode={543050},
city={Pamplona},
country={Colombia}}

\affiliation[2]{organization={Laboratorio de Simulación Control Biofotónica y Nanotecnología (SiCoBioNa), Science and Technology Department, Universidad Nacional de Quilmes, Roque Saenz Peña N◦ 352,},
city={Bernal 1876, Buenos Aires},
country={Argentina}}
\begin{abstract}
Protecting confidential information requires not only preventing unauthorized access to encrypted data but also concealing the very existence of the protected information. In this work, we propose a computational dual-layer optical security framework that integrates optical encryption and steganographic camouflage into a unified strategy. The proposed method is based on a 4F optical architecture and employs two two-dimensional private keys: a phase-only key represented through Circular Harmonic Components (CHC) and a periodic amplitude mask acting as a second secret key. Their combined action generates visually diverse steganograms from encrypted RGB images while preserving the correct recovery of the original information by authorized users. Unlike conventional optical encryption methods that produce easily recognizable cryptograms, the proposed approach disguises the encrypted information within camouflage patterns, providing an additional layer of protection before any decryption process is attempted. Numerical simulations demonstrate successful encryption, camouflage, and image recovery while showing that different steganographic appearances can be generated by modifying the private key and the periodic-mask parameters. Furthermore, a prospective optical implementation based on a Mach--Zehnder interferometer and digital holographic recording is presented, providing a feasible path toward future experimental realization. The proposed methodology is introduced as a proof of concept of the dual-layer optical security framework; a comprehensive cryptanalytic evaluation is beyond the scope of this first study and is left for future work.
\end{abstract}


\end{frontmatter}

\section{\textbf{Introduction}}

In this work, the term dual-layer optical security framework refers to an architecture in which the first security layer encrypts the information content, whereas the second layer conceals the existence of the encrypted information through steganographic camouflage.

Protecting confidential information during storage and transmission has become one of the fundamental challenges of the digital era. The rapid expansion of electronic communications, cloud computing, and distributed information systems has considerably increased the exposure of sensitive data to unauthorized access, interception, and cyberattacks. Consequently, the development of secure encryption techniques remains an active research field spanning classical cryptography, quantum cryptography, optical encryption, and information hiding strategies \cite{Danti2026-Stegano,EVERETT2024,HU2024,HAIDER2024,ZHANG2023,ref06_Subramani2023,Li2024a-Stegano,Li2024b-Stegano,Yu2023,Meng2024-Stegano,Verma2025-Stegano,Guerrero_V_Rueda_P_2021-Stegano,Rueda2015-CHC,Barrera2012-cripto,Chen2008-cripto,Chen2011-cripto,Clasica2020,Ding2014-cripto,Fraser2004-Seguridad,Ghoul2023-Stegano,Ito2024-Stegano,JASRA2022,Kunhoth2023-Stegano,QuantumvsClasica2021,Ravichandran2024-Stegano,Setiadi2023-Stegano,Saeidi2024-Stegano,Tong2024-Stegano,Javidi1996-crito,Javidi1999a-Cripto,Javidi1999b-4F,Joshi2010-cripto,L2016-cripto,Li2006-cripto,Li2010-cripto,Li2015-cripto,Lin2015-cripto,Liu2001-cripto,Ludia2006-Cripto,M1999-cripto,Meng2006-cripto,Muniraj2014-cripto,Nomura2000-Cripto,QuatumClasica2017,Refregier1995-cripto,Subramanian2021-Stegano,Tajahuerce2000-cripto,Tan2000-Cripto,Tebaldi2009-cripto,Valcourt1989,Wang2013-cripto,Wu2015-cripto}. Beyond ensuring the confidentiality of digital information, modern security systems increasingly seek to conceal the very existence of protected data, combining encryption and steganographic techniques to provide multiple layers of protection.

Optical cryptography has emerged as an attractive alternative to conventional digital encryption owing to its inherent parallelism, high processing speed, and capability for manipulating two-dimensional information. Since the pioneering work of Refregier and Javidi based on Double Random Phase Encoding (DRPE) implemented in a 4F optical processor, numerous optical encryption architectures have been proposed, including fractional Fourier transform methods, joint transform correlators, holographic encryption, diffractive optical systems, and polarization-based techniques. These approaches have demonstrated remarkable versatility for protecting optical information while taking advantage of the unique properties of coherent imaging systems. More recently, optical encryption has been combined with information hiding strategies to further increase the difficulty of detecting protected information and to enhance the overall security of optical communication systems.

Despite these significant advances, most optical encryption methods primarily concentrate on protecting the encrypted content while producing cryptograms that remain visually distinguishable from ordinary images. Consequently, although the encrypted information cannot be interpreted without the appropriate secret keys, its appearance may still reveal the existence of protected information and therefore attract unwanted attention. This observation motivates the development of optical security strategies that not only encrypt information but also disguise the visual appearance of the resulting cryptograms. Such an approach introduces an additional security layer by reducing the probability that encrypted information will be identified before any cryptanalytic attempt is performed.

Motivated by these challenges, this work proposes a computational dual-key optical encryption framework that integrates optical cryptography and steganographic camouflage into a unified security strategy. The proposed method combines a phase-only secret key represented through Circular Harmonic Components (CHC) with a second two-dimensional secret key consisting of a periodic amplitude mask. Their combined action enables the generation of visually diverse steganograms while preserving the successful recovery of the original encrypted information. Unlike conventional optical encryption schemes that generate easily recognizable cryptograms, the proposed approach conceals the encrypted information within camouflage patterns, thereby introducing a second layer of protection before any decryption process is attempted.

The main contributions of this work are summarized as follows:

\begin{enumerate}
    \item A dual-key optical encryption architecture based on a phase-only key represented through Circular Harmonic Components (CHC) and a periodic amplitude mask acting as a second private key.

    \item A cryptogram camouflage strategy capable of generating visually different steganograms without modifying the underlying encrypted information.

    \item A computational implementation demonstrating the encryption, camouflage, and successful recovery of RGB images using the proposed dual-key methodology.

    \item A prospective optical implementation based on a Mach--Zehnder interferometric configuration combined with digital holographic recording, providing a feasible path toward future experimental realization.
\end{enumerate}

The proposed methodology is presented as a proof of concept of a dual-layer optical security framework in which optical encryption protects the information content while steganographic camouflage conceals the existence of the cryptogram itself. A comprehensive cryptanalytic evaluation, including statistical security analyses, key sensitivity, robustness against perturbations, and resistance to cryptographic attacks, is beyond the scope of this first study and will be addressed in future work.

\section{\textbf{Method and Results}}
To establish the basis of the proposed methodology, this section first reviews the conventional optical encryption process implemented in a $4F$ architecture. This classical configuration provides the reference model from which the proposed dual-layer optical security framework is developed. After presenting the conventional encryption–decryption procedure and its computational implementation, the decomposition of the encryption key into Circular Harmonic Components (CHC) is introduced as the fundamental element that enables the proposed cryptogram camouflage strategy.

\subsection{\textbf{Conventional 4F Optical Encryption}}

The proposed methodology is built upon the classical Van der Lugt optical correlator architecture \cite{Lugt1964-4F}, one of the most widely employed configurations in optical encryption systems \cite{Javidi1996-crito,Javidi1999b-4F,Li2010-cripto}. This architecture, commonly referred to as the $4F$ optical processor, performs encryption and decryption by introducing a phase-only random mask in the Fourier plane. The phase mask acts as the private encryption key and is defined as

\begin{equation}
K(u,v)=e^{i\Phi(u,v)},
\end{equation}
is placed in the Fourier plane and acts as the encryption--decryption key. The function $\Phi(u,v)$ denotes a random phase distribution whose values are uniformly distributed over the interval $[-\pi,\pi]$, providing the randomness required for secure encryption.

\begin{figure}[H]
    \centering
    \begin{subfigure}[b]{0.6\textwidth}
        \centering
        \includegraphics[width=1\textwidth]{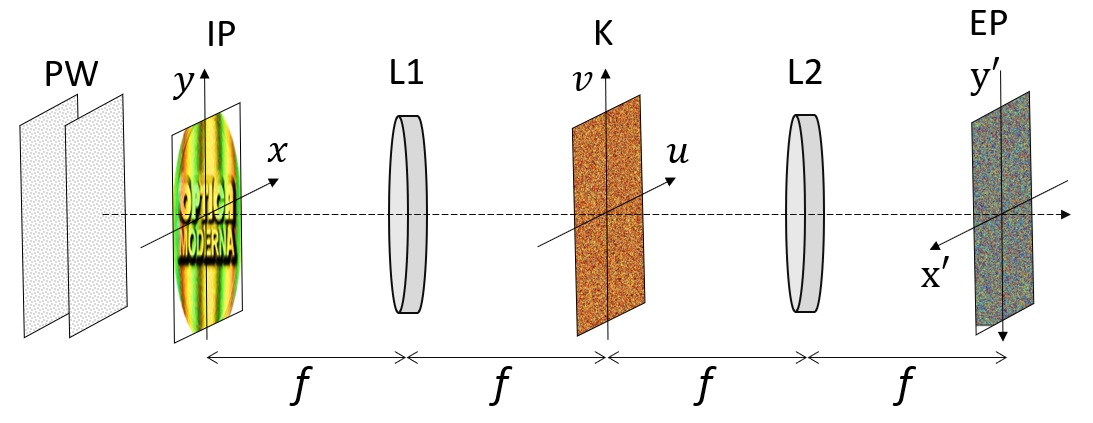} 
        \caption{4F encryption architecture.}
        \label{fig1a}
    \end{subfigure}
    
    \vfill
    \vspace{0.5cm}
    \begin{subfigure}[b]{0.6\textwidth}
        \centering
        \includegraphics[width=1\textwidth]{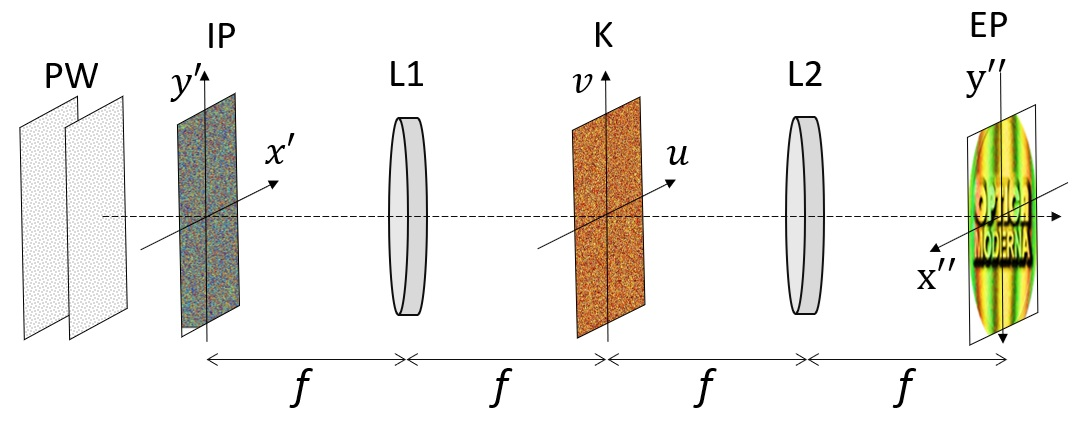} 
        \caption{4F decryption architecture.}
        \label{fig1b}
    \end{subfigure}
     \caption{4F optical architectures for encryption-decryption. PW: Plane wave; IP: Input plane; L1 and L2 are positive power lenses; f: focal length of each lens; K: Encryption-Decryption key; EP: Output plane.}
    \label{Fig1}
\end{figure}

When the phase mask is generated in rectangular coordinates, the encrypted image exhibits a white-noise-like appearance. Consequently, different encrypted images become visually indistinguishable, making direct interpretation of the protected information impossible. Nevertheless, the resulting cryptograms still possess the characteristic appearance of encrypted data, allowing an observer to immediately recognize that protected information is being transmitted. Figure~\ref{Fig1} illustrates the conventional optical encryption and decryption architecture employed as the reference system throughout this work.

\subsubsection*{Encryption}

The secret image, $f(x,y)$, is located at the input plane and illuminated by a monochromatic plane wave. Lens $L_1$ computes the Fourier transform of the input image, producing the spectrum $F(u,v)$ at its back focal plane \cite{Goodman1996}. This spectrum is multiplied by the phase-only encryption key

\begin{equation}
K_e(u,v)=e^{i\Phi(u,v)},
\end{equation}

where $\Phi(u,v)$ is a uniformly distributed random phase function over the interval $[-\pi,\pi]$.

The inverse Fourier transform performed by lens $L_2$ produces the encrypted image

\begin{equation}
f_e(x',y')
=
f(x',y')
*
k_e(x',y'),
\label{eq:encryption}
\end{equation}

where $k_e(x,y)$ is the Fourier transform of $K_e(u,v)$ and $*$ denotes convolution.

Since $k_e(x,y)$ corresponds to a random phase distribution, the encrypted image exhibits a white-noise-like appearance, effectively concealing the visual information contained in the original image.

\subsubsection*{Decryption}

For decryption, the complex conjugate of the encrypted image, $f_e^{*}(x',y')$, is placed at the input plane of the second $4F$ processor. After Fourier transformation, the spectrum

\begin{equation}
F_e^{*}(-u,-v)
\end{equation}

is multiplied by the decryption key $K_d(u,v)$.

The inverse Fourier transform performed by lens $L_2$ yields

\begin{equation}
f_d(x'',y'')
=
f(x'',y'')
*
k_e^{*}(x'',y'')
\circledast
k_d(x'',y''),
\label{eq:decryption}
\end{equation}

where $\circledast$ denotes correlation.

When the decryption key is identical to the encryption key,

\begin{equation}
k_d(x,y)=k_e(x,y),
\end{equation}

their correlation becomes

\begin{equation}
k_e^{*}(x,y)\circledast k_e(x,y)=\delta(x,y),
\end{equation}

which leads to perfect recovery of the original image,

\begin{equation}
f_d(x,y)=f(x,y).
\end{equation}

Conversely, if $k_d(x,y)\neq k_e(x,y)$, the output consists of a random noise distribution, making reconstruction of the secret image impossible.

The security of the proposed architecture can be further enhanced by introducing an additional phase-only random mask at the input plane, thereby increasing the effective key space and the resistance against unauthorized decryption.

\begin{figure}[H]
\centering
\includegraphics[width=10cm]{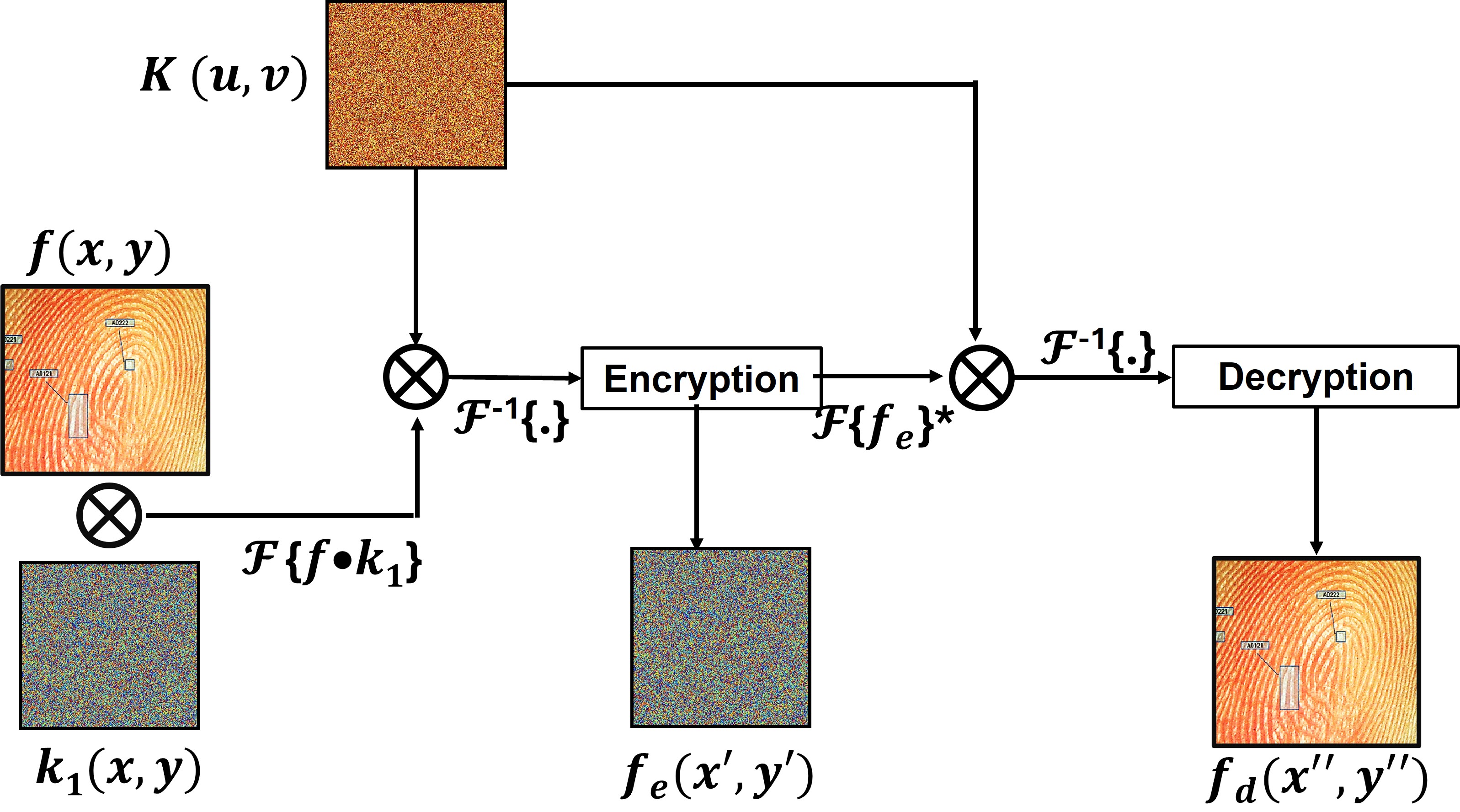} 
\caption{Flowchart of the conventional double-mask optical encryption--decryption algorithm based on the $4F$ architecture. The symbol $\otimes$ denotes pointwise multiplication. $\mathcal{F}\{\cdot\}$, $\mathcal{F}^{-1}\{\cdot\}$, and $\mathcal{F}\{\cdot\}^{*}$ denote the Fourier transform, inverse Fourier transform, and complex-conjugate Fourier transform, respectively.}
\label{fig2}
\end{figure}

Figure~\ref{fig2} summarizes the implementation of the conventional double-mask optical encryption--decryption algorithm based on the $4F$ architecture of Fig.~\ref{Fig1}. Both the encryption and decryption keys are phase-only masks generated in rectangular coordinates.

\begin{figure}[H]
    \centering
    \begin{subfigure}[b]{0.45\textwidth}
        \centering
        \includegraphics[width=0.7\textwidth]{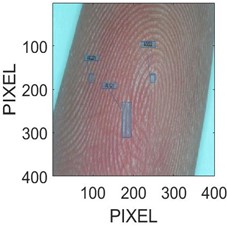} 
        \caption{Input image.}
        \label{fig3a}
    \end{subfigure}
    \hspace{0.1cm}
    \begin{subfigure}[b]{0.45\textwidth}
        \centering
        \includegraphics[width=0.7\textwidth]{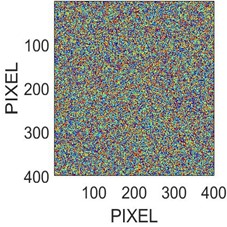} 
        \caption{Phase key.}
        \label{fig3b}
    \end{subfigure}
    \vfill
    \vspace{0.5cm}
    \begin{subfigure}[b]{0.45\textwidth}
        \centering
        \includegraphics[width=0.7\textwidth]{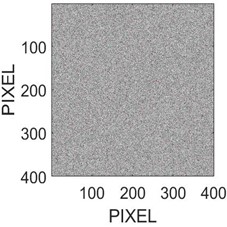} 
        \caption{Amplitude of the encrypted image.}
        \label{fig3c}
    \end{subfigure}
    \hspace{0.1cm}
    \begin{subfigure}[b]{0.45\textwidth}
        \centering
        \includegraphics[width=0.7\textwidth]{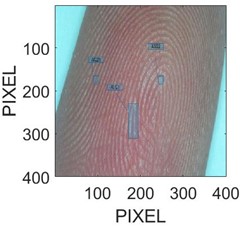} 
        \caption{Decrypted image.}
        \label{fig3d}
    \end{subfigure}
    \caption{Computational result using a double phase-only mask in rectangular coordinates.}
    \label{Fig3}
\end{figure}

To establish a reference for evaluating the proposed method, the conventional algorithm was implemented in MATLAB. Figure~\ref{Fig3} presents a representative cryptogram obtained with this implementation, where both phase masks were generated in rectangular coordinates. The resulting cryptograms serve as the benchmark for comparison with the proposed steganographic cryptogram camouflage approach.

Figure~\ref{Fig3} presents a representative result obtained with the conventional double-mask optical encryption algorithm shown in Fig.~\ref{fig2}. The encryption and decryption phase masks were generated in rectangular coordinates. The original image is shown in Fig.~\ref{fig3a}, the corresponding encrypted image (cryptogram) in Fig.~\ref{fig3c}, and the decrypted image in Fig.~\ref{fig3d}.

Visual inspection indicates that the decrypted image closely matches the original scene. However, evaluation of the absolute difference,

\begin{equation}
\left|f_d(x,y)-f(x,y)\right|,
\end{equation}

reveals the presence of a small residual error caused by numerical noise introduced during the encryption--decryption process. To quantitatively assess the reconstruction quality, the Peak Signal-to-Noise Ratio (PSNR) was computed for the decrypted image, yielding a value of 52.46~dB. This value indicates excellent reconstruction quality, with differences between the original and recovered images being practically imperceptible to the human eye.

Although the conventional double-mask architecture provides excellent reconstruction quality, all encrypted images exhibit nearly identical white-noise characteristics regardless of the protected content. Consequently, while the information remains cryptographically protected, the visual appearance of the cryptograms still reveals that encrypted information is being transmitted. Therefore, the objective is not to improve the reconstruction quality of the conventional 4F encryption system, but to transform the visual appearance of the cryptograms into steganographic camouflage patterns while preserving the correct recovery of the original image. To achieve this objective, the conventional rectangular-coordinate representation of the phase key is replaced by a representation based on Circular Harmonic Components (CHC), as described in the following section.

\subsection{\textbf{Decomposition of the Encryption Key into Circular Harmonic Components}}

The previous section demonstrated that the conventional $4F$ optical encryption architecture provides excellent reconstruction quality but generates cryptograms with nearly identical white-noise characteristics. Although this random appearance effectively conceals the image content, it also reveals that encrypted information is being transmitted. To overcome this limitation, the proposed methodology replaces the conventional phase-key representation in rectangular coordinates with a representation based on Circular Harmonic Components (CHCs). This alternative representation constitutes the foundation of the proposed cryptogram camouflage strategy while preserving the random phase properties required for optical encryption.

The proposed CHC decomposition was implemented in MATLAB following the algorithm introduced by Gualdrón and Arsenault \cite{Gualdron1993-CHC}, originally developed for the synthesis of rotation-invariant correlation filters. In the present work, this formulation is adapted to generate phase-only encryption keys suitable for the proposed dual-layer optical security framework.

As an example, this algorithm was implemented using Matlab to compare with the results of our new proposal of camouflaging the cryptogram in a steganogram.

Let $f(x,y)$ denote an image defined in Cartesian coordinates. To construct the proposed encryption key, the image is first represented in polar coordinates, $f(r,\theta)$, and subsequently decomposed into its Circular Harmonic Components (CHCs) using the Fourier series expansion described in Refs.~\cite{Rueda2015-CHC,Hansen1981-CHC,Hansen1981a-CHC,Gualdron1993-CHC},

\begin{equation}
f_{\mathrm{CHC}}(r,\theta)
=
\sum_{m=-\infty}^{\infty}
f_m(r)e^{im\theta},
\label{eq:chc}
\end{equation}

where the circular harmonic coefficients are given by

\begin{equation}
f_m(r)
=
\frac{1}{2\pi}
\int_{0}^{2\pi}
f(r,\theta)e^{-im\theta}\,d\theta.
\label{eq:fm}
\end{equation}

Here, $r$ and $\theta$ denote the radial and angular coordinates, respectively, while $m$ is the integer order of the circular harmonic expansion. Equation~(\ref{eq:fm}) is evaluated after transforming the original image from Cartesian coordinates, $f(x,y)$, to its polar representation, $f(r,\theta)$, as illustrated in Fig.~\ref{Fig4}.

To illustrate the decomposition procedure, a synthetic rectangular ring was selected as a representative test object. Each image sample is first mapped from Cartesian coordinates to its corresponding position in the polar domain before computing the Circular Harmonic Components according to Eqs.~(\ref{eq:chc}) and (\ref{eq:fm}).

Figure~\ref{Fig5} presents a representative CHC decomposition obtained with the implemented algorithm. Figures~\ref{fig5b}--\ref{fig5d} show the phase, real part, and magnitude distributions of the circular harmonic decomposition, respectively.

\section*{\textbf{Algorithm for Circular Harmonic Components Decomposition:}}

\begin{figure}[H]
    \centering
    \begin{subfigure}[b]{0.45\textwidth}
        \centering
        \includegraphics[width=0.7\textwidth]{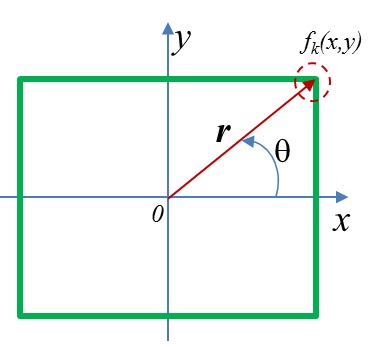} 
        \caption{$f(x, y)$.}
        \label{fig4a}
    \end{subfigure}
    \hspace{0.1cm}
    \begin{subfigure}[b]{0.45\textwidth}
        \centering
        \includegraphics[width=0.9\textwidth]{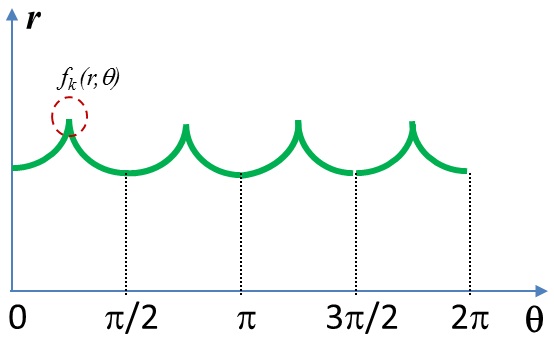} 
        \caption{$f(r,\theta)$.}
        \label{fig4b}
    \end{subfigure}
    
    \vspace{0.5cm}
    
    \begin{subfigure}[b]{0.45\textwidth}
        \centering
        \includegraphics[width=0.68\textwidth]{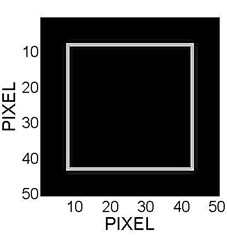} 
        \caption{Image in Cartesian coordinates $f(x,y)$.}
        \label{fig4c}
    \end{subfigure}
    \hspace{0.1cm}
    \begin{subfigure}[b]{0.45\textwidth}
        \centering
        \includegraphics[width=0.9\textwidth]{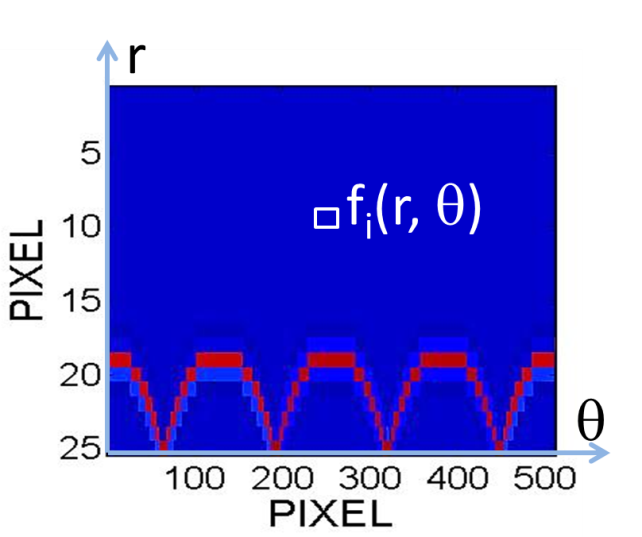} 
        \caption{Image in polar coordinates $f(r,\theta)$.}
        \label{fig4d}
    \end{subfigure}
    
    \caption{Transformation of the input image from Cartesian coordinates, $f(x,y)$, to its polar representation, $f(r,\theta)$.}
    \label{Fig4}
\end{figure}

The CHC decomposition algorithm was implemented in MATLAB. To illustrate its operation, a synthetic $50\times50$ pixel image containing a square ring, shown in Fig.~\ref{fig4c}, was used as a test pattern. The first stage of the algorithm transforms the image from Cartesian coordinates, $f(x,y)$, to its polar representation, $f(r,\theta)$.

The resulting harmonic representation provides an alternative description of the phase key while preserving its statistical randomness. This property is essential because it allows the encryption key to be reformulated without compromising its suitability for optical encryption.

The input image is represented as an $M\times M$ matrix, whose origin is initially located at the upper-left corner. Before performing the coordinate transformation, the origin is translated to the center of the image, $(M/2,M/2)$, thereby establishing the reference point for the polar mapping. The image is then sampled along the radial direction, as illustrated in Fig.~\ref{fig4a}, and subsequently represented in the polar domain.

The polar representation is obtained by sampling the image along the angular direction over the interval $0\leq\theta<2\pi$, producing the matrix $f(r,\theta)$. The resulting polar image is shown in Fig.~\ref{fig4d}. In this representation, the image columns correspond to uniformly sampled angular positions, whereas the rows represent the radial coordinate.

Once the polar image has been computed, the circular harmonic coefficients are evaluated using the discrete form of Eq.~(\ref{eq:fm}),

\begin{equation}
f_m(r)=
\frac{1}{n}
\sum_{k=1}^{n}
f\!\left(r,\frac{2\pi k}{n}\right)
\exp\!\left(-im\frac{2\pi k}{n}\right),
\label{eq:discreteCHC}
\end{equation}

where $n$ is the number of uniformly spaced angular samples over the interval $[0,2\pi)$. After computing the coefficients for all harmonic orders, the Circular Harmonic Expansion is reconstructed according to Eq.~(\ref{eq:chc}), yielding the matrix $f_{\mathrm{CHC}}(r,\theta)$.

\begin{figure}[H]
    \centering
    \begin{subfigure}[b]{0.45\textwidth}
        \centering
        \includegraphics[width=0.6\textwidth]{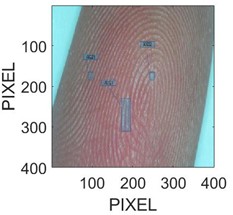} 
        \caption{Image of $f(x, y)$.}
        \label{fig5a}
    \end{subfigure}
    \hspace{0.1cm}
    \begin{subfigure}[b]{0.45\textwidth}
        \centering
        \includegraphics[width=0.6\textwidth]{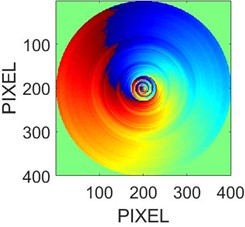} 
        \caption{Decomposition phase.}
        \label{fig5b}
    \end{subfigure}
    
    \vspace{0.5cm}
    
    \begin{subfigure}[b]{0.45\textwidth}
        \centering
        \includegraphics[width=0.6\textwidth]{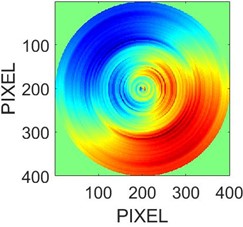} 
        \caption{Real part of the decomposition.}
        \label{fig5c}
    \end{subfigure}
    \hspace{0.1cm}
    \begin{subfigure}[b]{0.45\textwidth}
        \centering
        \includegraphics[width=0.6\textwidth]{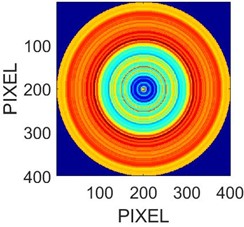} 
        \caption{Module of the decomposition.}
        \label{fig5d}
    \end{subfigure}
        \caption{Circular Harmonic Components (order$m=1$) decomposition.}
    \label{Fig5}
\end{figure}
Figure~\ref{Fig5} presents the CHC decomposition obtained from the test image shown in Fig.~\ref{fig4c}. For visualization purposes, the resulting matrix is shifted so that its origin is located at the center of the image, $(M/2,N/2)$.

\begin{figure}[H]
    \centering
    \begin{subfigure}[b]{0.45\textwidth}
        \centering
        \includegraphics[width=0.6\textwidth]{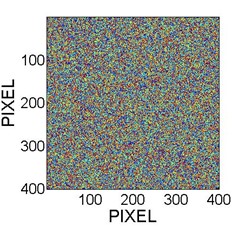} 
        \caption{Key phase $K(u,v)$.}
        \label{fig6a}
    \end{subfigure}
    \hspace{0.1cm}
    \begin{subfigure}[b]{0.45\textwidth}
        \centering
        \includegraphics[width=0.6\textwidth]{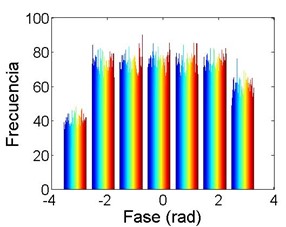} 
        \caption{Histogram of (a).}
        \label{fig6b}
    \end{subfigure}
    
    \vspace{0.5cm}
    
    \begin{subfigure}[b]{0.45\textwidth}
        \centering
        \includegraphics[width=0.6\textwidth]{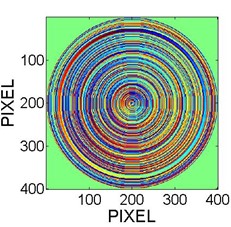} 
        \caption{Key phase $K_m (\rho,\phi)$ for $m=1$.}
        \label{fig6c}
    \end{subfigure}
    \hspace{0.1cm}
    \begin{subfigure}[b]{0.45\textwidth}
        \centering
        \includegraphics[width=0.6\textwidth]{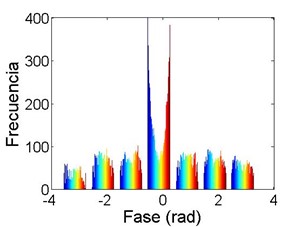} 
        \caption{Histogram of (c).}
        \label{fig6d}
    \end{subfigure}
    
    \caption{Key in Cartesian coordinates and its CHC decomposition.}
    \label{Fig6}
\end{figure}

Figure~\ref{Fig6} compares the random phase distributions generated in Cartesian coordinates with their corresponding representations obtained from the Circular Harmonic Expansion. The associated phase histograms, shown in Figs.~\ref{fig6b} and \ref{fig6d}, exhibit an approximately uniform distribution over the interval $[-\pi,\pi]$. This result indicates that both key representations preserve the random phase characteristics required for optical encryption, making them suitable for use as phase-only encryption keys.

Therefore, the proposed Circular Harmonic representation preserves the statistical phase properties of conventional random phase masks while providing an alternative key representation for the proposed encryption framework.

The comparison shown in Fig.~\ref{Fig6} demonstrates that the Circular Harmonic representation preserves the approximately uniform phase distribution exhibited by conventional random phase masks. Consequently, the proposed CHC-based keys maintain the statistical properties required for optical encryption while providing a structured spatial representation that will subsequently be exploited to generate visually distinguishable cryptograms. This characteristic constitutes the key element that enables the proposed cryptogram camouflage strategy.

\begin{figure}[H]
\centering
\includegraphics[width=10cm]{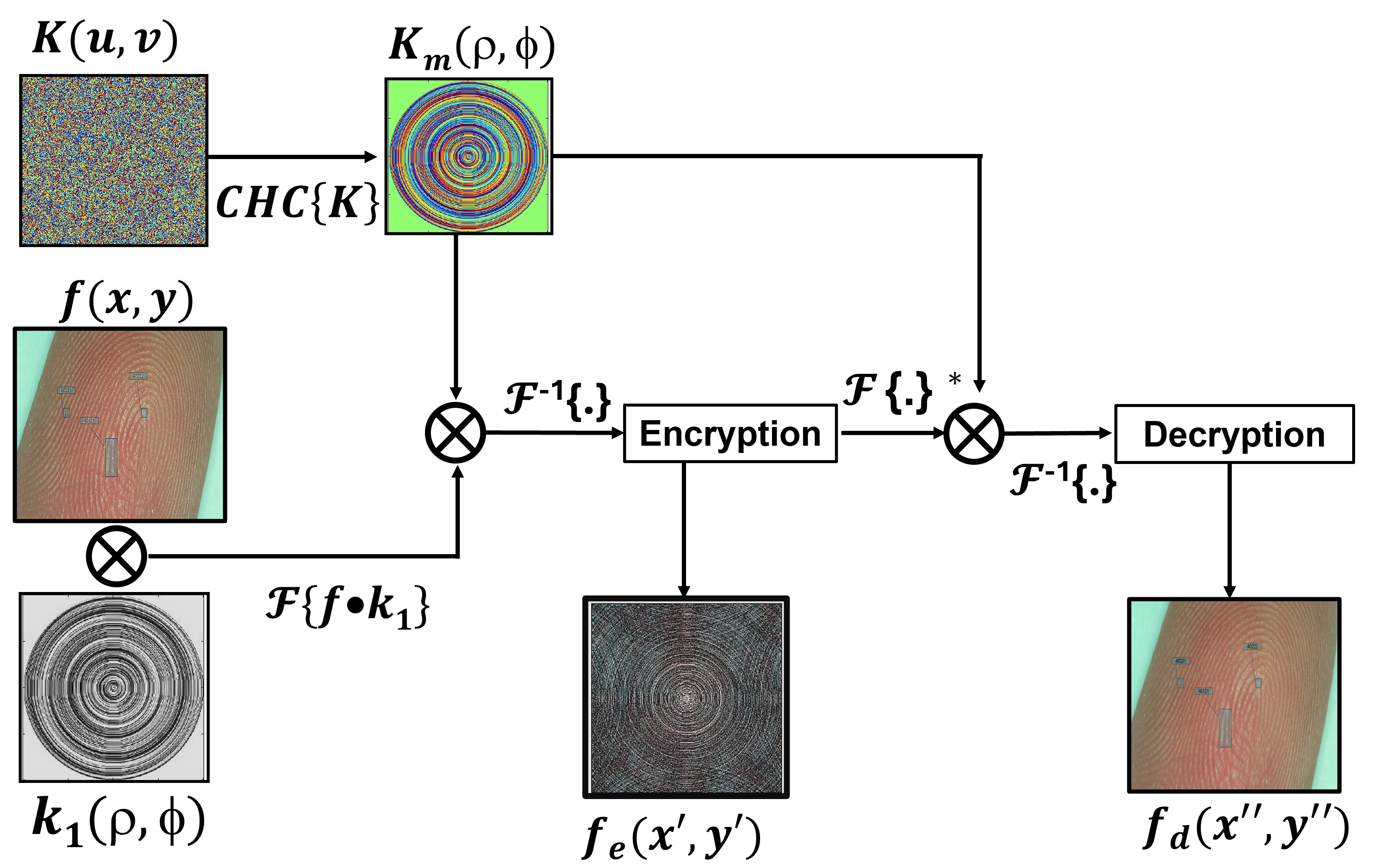} 
\caption{Flowchart of the proposed optical encryption--decryption algorithm using a double private key represented by Circular Harmonic Components (CHCs). The symbol $\otimes$ denotes pointwise multiplication. $\mathcal{F}\{\cdot\}$, $\mathcal{F}^{-1}\{\cdot\}$, and $\mathcal{F}\{\cdot\}^{*}$ denote the Fourier transform, inverse Fourier transform, and complex-conjugate Fourier transform, respectively.}
\label{fig7}
\end{figure}

Once the phase key has been represented through Circular Harmonic Components, the conventional encryption procedure can be reformulated using the CHC-based key. Figure~\ref{fig7} summarizes the complete computational implementation of the proposed encryption–decryption algorithm employing the dual-key strategy.

\begin{figure}[H]
    \centering
    \begin{subfigure}[b]{0.45\textwidth}
        \centering
        \includegraphics[width=0.6\textwidth]{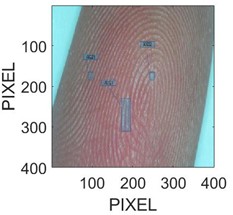} 
        \caption{Input image.}
        \label{fig8a}
    \end{subfigure}
    \hspace{0.1cm}
    \begin{subfigure}[b]{0.45\textwidth}
        \centering
        \includegraphics[width=0.6\textwidth]{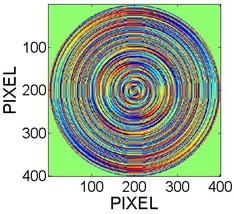} 
        \caption{Key phase $K_m (\rho,\phi)$ for $m=1$.}
        \label{fig8b}
    \end{subfigure}
        \vfill
        \begin{subfigure}[b]{0.45\textwidth}
        \centering
        \includegraphics[width=0.6\textwidth]{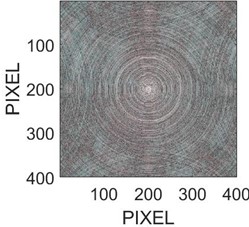} 
        \caption{Amplitude of the encrypted image.}
        \label{fig8c}
    \end{subfigure}
    \hspace{0.1cm}
    \begin{subfigure}[b]{0.45\textwidth}
        \centering
        \includegraphics[width=0.6\textwidth]{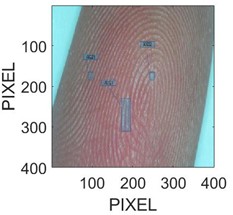} 
        \caption{Decrypted image.}
        \label{fig8d}
    \end{subfigure}
        \caption{Representative encryption and decryption results obtained with the proposed CHC-based optical encryption algorithm. (a) Original RGB image. (b) Phase distribution of the CHC encryption key corresponding to the harmonic order $m=1$. (c) Amplitude of the encrypted complex image (cryptogram). (d) Decrypted image. The recovered image exhibits a PSNR of 41.19~dB.}
    \label{Fig8}
\end{figure}

A representative encryption result obtained with the proposed approach is shown in Fig.~\ref{Fig8}. The input consists of a $400\times400$ pixel RGB image. Additional experiments were performed using different types of input scenes, yielding similar encryption and decryption performance. Consequently, the characteristics of the proposed method were found to be essentially independent of the image content.

Unlike the conventional implementation, whose cryptograms exhibit an almost featureless white-noise appearance, the proposed CHC-based representation generates encrypted images with well-defined circular structures determined by the harmonic decomposition. These structured cryptograms constitute an intermediate encrypted representation rather than the final protected image. In the following section, these characteristic patterns are further transformed into visually innocuous steganograms through the proposed periodic-amplitude masking strategy, completing the second layer of the proposed optical security framework.

\section{\textbf{Steganographic Camouflage of the Cryptogram}}

The previous section demonstrated that the proposed CHC-based encryption key generates structured cryptograms whose appearance differs substantially from the white-noise patterns produced by conventional optical encryption. Building upon this property, the present section introduces the second layer of the proposed optical security framework, in which the cryptogram is transformed into a visually innocuous steganogram through periodic spatial modulation. Consequently, the transmitted image conceals not only the information content but also the existence of the encrypted message itself.

Steganography aims to conceal the existence of a secret message by embedding it into an innocuous carrier, thereby preventing its detection by unauthorized observers \cite{Li2024b-Stegano,Gan2024-Stegano,Ito2024-Stegano,Meng2024-Stegano,Tong2024-Stegano,Danti2026-Stegano,Ghoul2023-Stegano,Verma2025-Stegano,Saeidi2024-Stegano,Kunhoth2023-Stegano,Setiadi2023-Stegano,Subramanian2021-Stegano,Ravichandran2024-Stegano,Guerrero_V_Rueda_P_2021-Stegano} In contrast, cryptography protects the message by transforming it into an unintelligible form that can only be recovered with the appropriate decryption key.

The proposed method combines both approaches. Rather than transmitting the optical cryptogram directly, the encrypted image is first camouflaged within a host image to generate a steganogram. Consequently, the existence of the encrypted message is concealed while preserving the security provided by the optical encryption process.

Unlike conventional steganographic methods, where a secret message is embedded into a host image, the proposed approach camouflages the optical cryptogram itself before transmission. The resulting steganogram preserves the security provided by the optical encryption process while reducing the probability that the encrypted information will be visually identified.

The proposed steganographic camouflage method is based on modulating the input image with an inverse periodic mask, denoted by $h^{-1}(ax_n,by_n)$, where $a$, $b$, and $n$ are design parameters that determine the periodic structure of the mask.

\begin{figure}
\centering
\includegraphics[width=14cm]{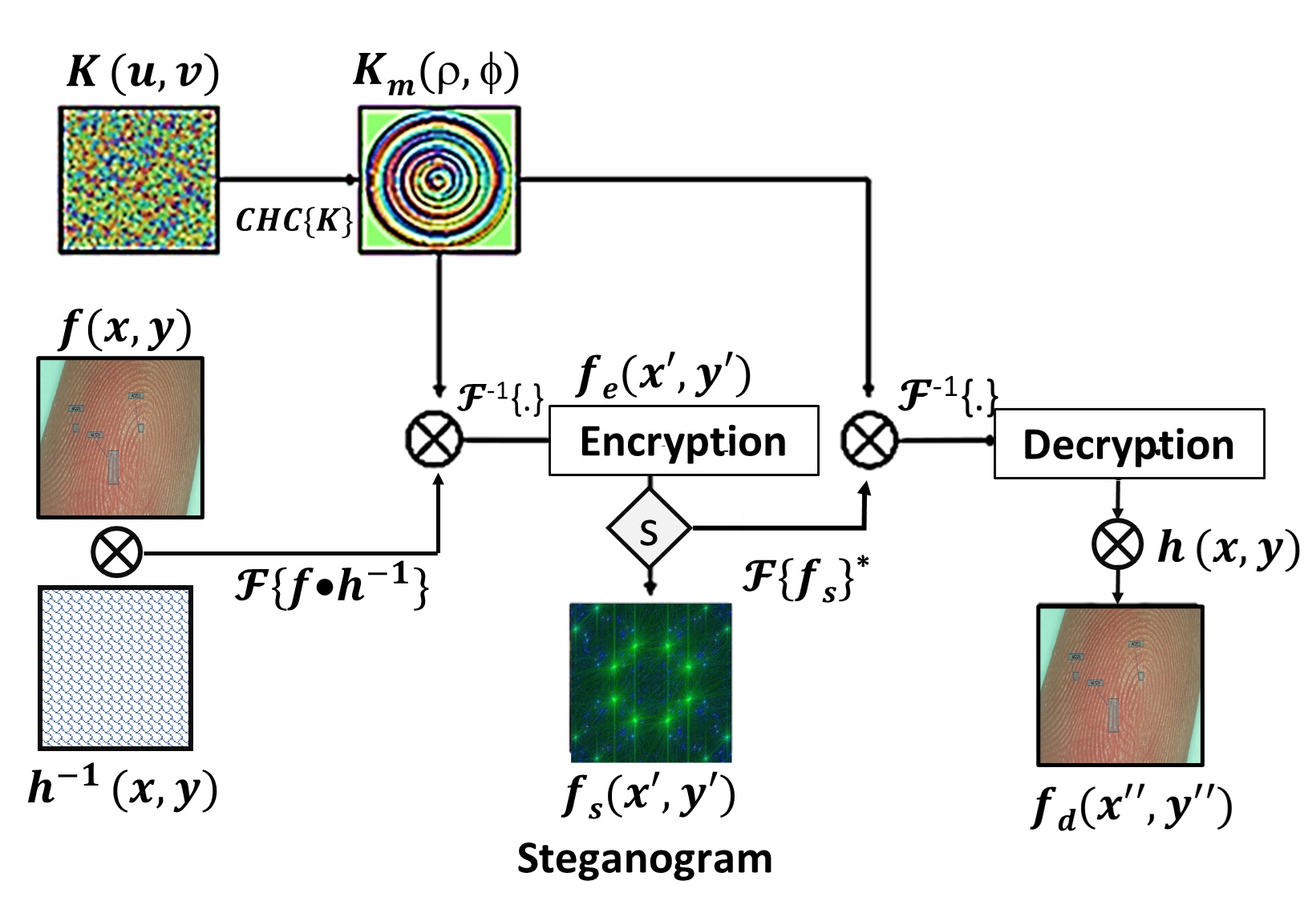} 
\caption{Flowchart of the proposed optical encryption, steganographic camouflage, and decryption algorithm using double CHC-based private keys. The symbol $\otimes$ denotes pointwise multiplication. $\mathcal{F}\{\cdot\}$, $\mathcal{F}^{-1}\{\cdot\}$, and $\mathcal{F}\{\cdot\}^{*}$ denote the Fourier transform, inverse Fourier transform, and complex-conjugate Fourier transform, respectively.}
\label{fig9}
\end{figure}

Figure~\ref{fig9} summarizes the complete computational workflow of the proposed dual-layer optical security framework. The first stage performs optical encryption using the CHC-based private key, whereas the second stage applies periodic spatial modulation to transform the resulting cryptogram into a steganographic camouflage pattern. During decryption, the inverse operations recover the original encrypted information only when both private keys are correctly provided.

In the proposed methodology, the cryptogram is no longer regarded as the final transmitted object. Instead, it constitutes an intermediate encrypted representation that is subsequently transformed into a family of visually distinct steganograms through a second private key based on periodic spatial modulation. Consequently, the transmitted information conceals not only the image content but also the existence of the encrypted message itself, introducing the second layer of the proposed optical security framework.

Mathematically, the steganographic stage can be interpreted as an additional modulation process applied to the encrypted image before transmission. The resulting steganogram is expressed as
\begin{equation}
f_s(x',y')
=
\mathcal{F}^{-1}\{G(u,v)\}
*
\mathcal{F}\{K_m(\rho,\phi)\},
\label{eq:stego}
\end{equation}

where

\begin{equation}
G(u,v)
=
\left[
F(u,v)*H(u,v)
\right]
,
\label{eq:Guv}
\end{equation}

and $F(u,v)$ and $H(u,v)$ denote the Fourier transforms of the secret image $f(x,y)$ and the inverse periodic mask $h^{-1}(x,y)$, respectively.

The inverse periodic mask introduces a deterministic spatial modulation that controls the visual appearance of the transmitted image, whereas the CHC-based phase key preserves the cryptographic functionality of the optical encryption process. Consequently, the final steganogram is jointly determined by two independent private keys: the CHC phase distribution and the periodic amplitude mask. This dual-key configuration constitutes the second layer of the proposed security framework.

The periodic mask introduces the spatial modulation required to camouflage the encrypted image, whereas the CHC-based encryption key preserves the security of the optical encryption process. Consequently, the visual appearance of the resulting steganogram is jointly determined by the periodic mask and the circular harmonic representation of the encryption key.

The decryption process recovers the secret image according to

\begin{equation}
f_d(x'',y'')
=
\mathcal{F}^{-1}
\left\{
\mathcal{F}\left\{f_s(x',y')\right\}^{*}
K_m(\rho,\phi)
\right\}
\,h(x'',y''),
\label{eq:decryption_stego}
\end{equation}

where $h(x,y)$ is the periodic mask associated with the inverse modulation mask employed during the encryption stage. Successful recovery of the secret image requires the correct CHC-based private key together with the corresponding periodic mask parameters.

To facilitate the validation of the proposed methodology, a dedicated graphical user interface (GUI) was developed. The interface integrates all stages of the encryption, camouflage, and decryption processes, allowing the user to configure the CHC order, the periodic-mask parameters, and the optical encryption keys within a unified computational environment.

\begin{figure}[H]
    \centering
    \begin{subfigure}[b]{1\textwidth}
        \centering
        \includegraphics[width=1\textwidth]{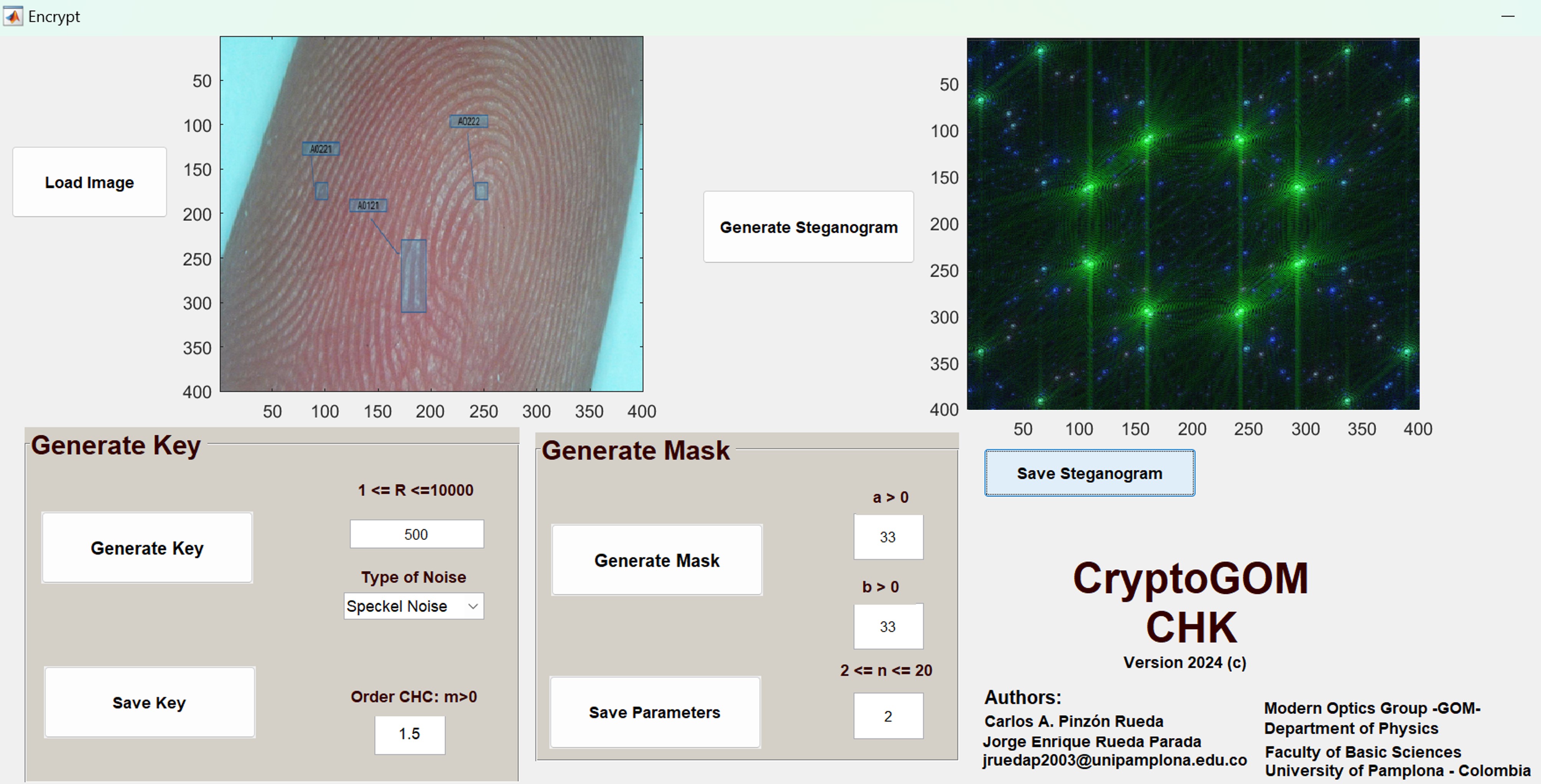} 
        \caption{GUI $\ Encryption \rightarrow Steganogram$.}
        \label{fig10a}
    \end{subfigure}
    
    \vfill
    \vspace{0.5cm}
    \begin{subfigure}[b]{1\textwidth}
        \centering
        \includegraphics[width=1\textwidth]{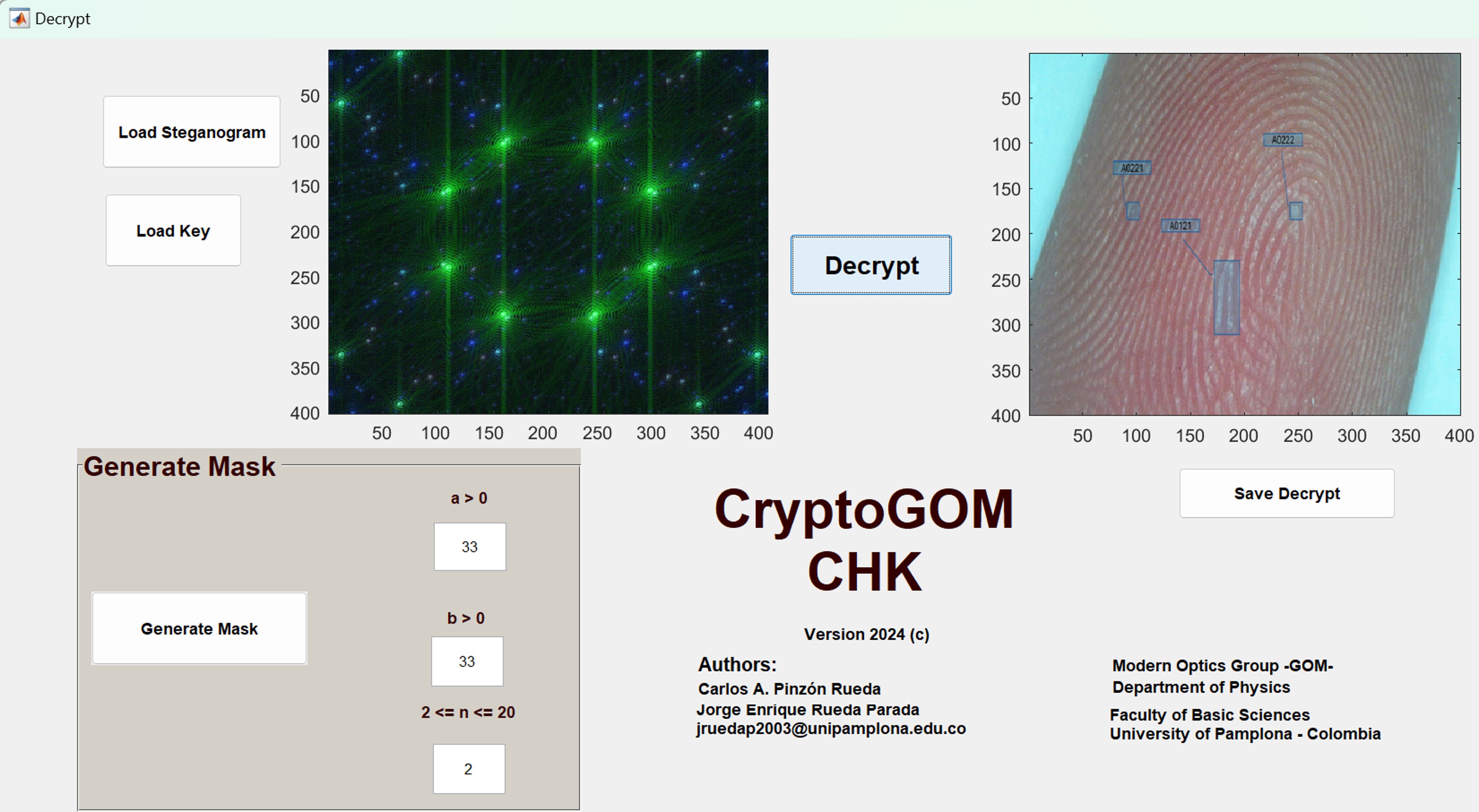} 
        \caption{GUI $Steganogram \rightarrow Decryption$.}
        \label{fig10b}
    \end{subfigure}
     \caption{GUI of the developed computational system.}
    \label{Fig10}
\end{figure}

Figure~\ref{Fig10} presents the graphical user interface (GUI) developed for the proposed encryption--decryption system. During encryption [Fig.~\ref{fig10a}], the user loads the secret image and specifies the system parameters, including the noise level, noise type, CHC order, and the mask parameters $(a,b,n)$ defining the inverse periodic mask $h^{-1}(ax^n,by^n)$. The CHC-based private key and the corresponding inverse periodic mask are then generated and stored, together with the associated parameters $(a,b,n)$. Finally, the steganographic cryptogram is generated and saved.


For decryption [Fig.~\ref{fig10b}], the recipient loads the steganogram, the corresponding CHC-based private key, and the private parameters $(a,b,n)$ required to reconstruct the periodic mask. The original image is recovered only when the correct key and mask parameters are employed.

\begin{figure}[H]
    \centering
    \begin{subfigure}[b]{0.45\textwidth}
        \centering
        \includegraphics[width=0.6\textwidth]{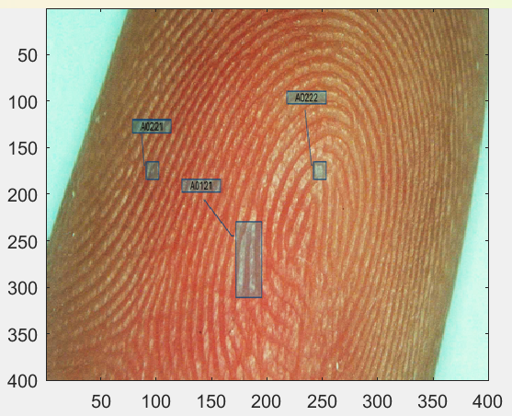} 
        \caption{Input image}
        \label{fig11a}
    \end{subfigure}
    \hspace{0.1cm}
    \begin{subfigure}[b]{0.45\textwidth}
        \centering
        \includegraphics[width=0.6\textwidth]{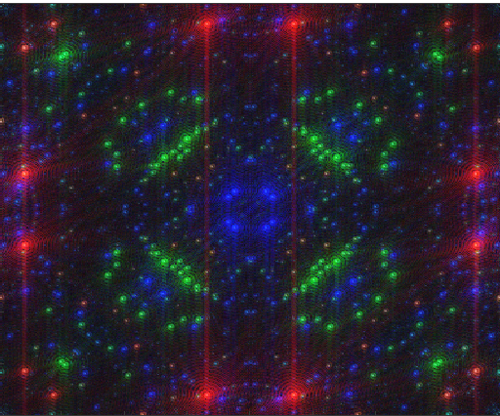} 
        \caption{$a=b=33; n=6; m=1.5$}
        \label{fig11b}
    \end{subfigure}
    \vfill
    \vspace{0.5cm}
    \begin{subfigure}[b]{0.45\textwidth}
        \centering
        \includegraphics[width=0.6\textwidth]{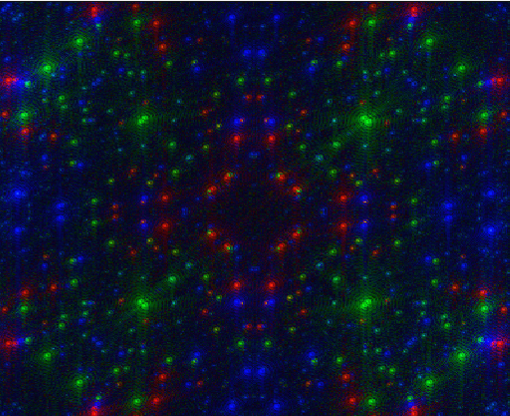} 
        \caption{$a=b=33 ; n=4; m=1.5$.}
        \label{fig11c}
    \end{subfigure}
    \hspace{0.1cm}
    \begin{subfigure}[b]{0.45\textwidth}
        \centering
        \includegraphics[width=0.6\textwidth]{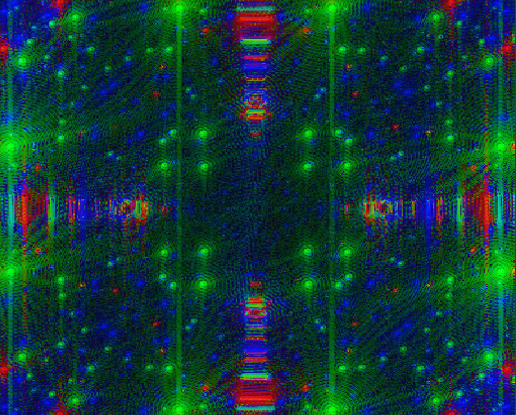} 
        \caption{$a=b=33; n=14; m=1.5$.}
        \label{fig11d}
    \end{subfigure}
     \vfill
    \vspace{0.5cm}
    \begin{subfigure}[b]{0.45\textwidth}
        \centering
        \includegraphics[width=0.6\textwidth]{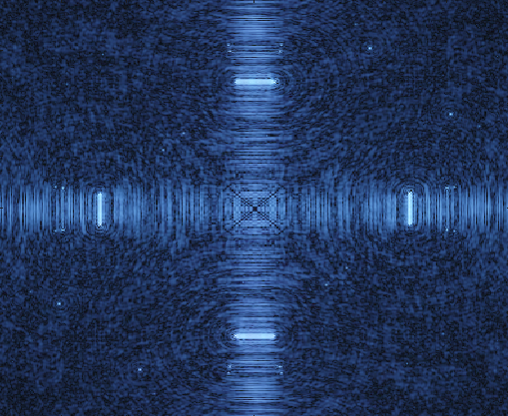} 
        \caption{$a=b=15; n=19; m=1.5$.}
        \label{fig11e}
    \end{subfigure}
    \hspace{0.1cm}
    \begin{subfigure}[b]{0.45\textwidth}
        \centering
        \includegraphics[width=0.6\textwidth]{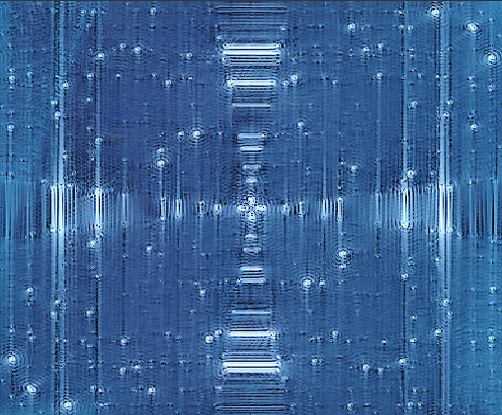} 
        \caption{$a=b=15; n=19; m=1.5$.}
        \label{fig11f}
    \end{subfigure}
    \caption{Representative steganograms generated with the proposed CHC-based optical encryption method. (a) Original image. (b)--(f) Steganograms obtained using different combinations of the private parameters $(a,b,n)$ and the CHC order $m$. The diversity of the resulting camouflage patterns demonstrates the flexibility of the proposed steganographic scheme.}
    \label{Fig11}
\end{figure}

\begin{figure}[H]
    \centering
    \begin{subfigure}[b]{0.45\textwidth}
        \centering
        \includegraphics[width=0.6\textwidth]{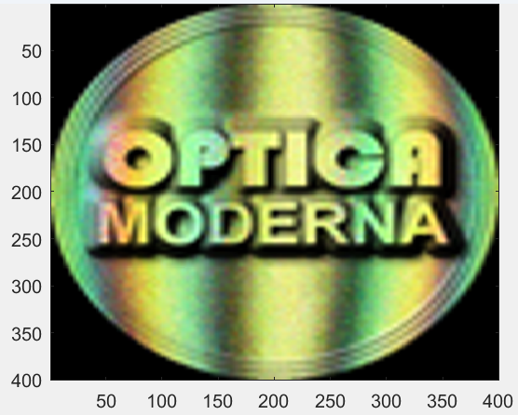} 
        \caption{Input image}
        \label{fig12a}
    \end{subfigure}
    \hspace{0.1cm}
    \begin{subfigure}[b]{0.45\textwidth}
        \centering
        \includegraphics[width=0.6\textwidth]{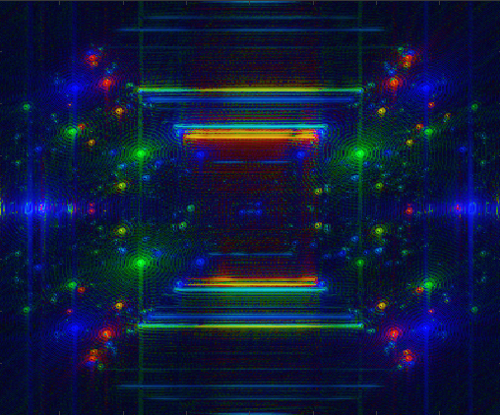} 
        \caption{$a=3; b=15; n=20; m=1.5$.}
        \label{fig12b}
    \end{subfigure}
    \vfill
    \vspace{0.5cm}
    \begin{subfigure}[b]{0.45\textwidth}
        \centering
        \includegraphics[width=0.6\textwidth]{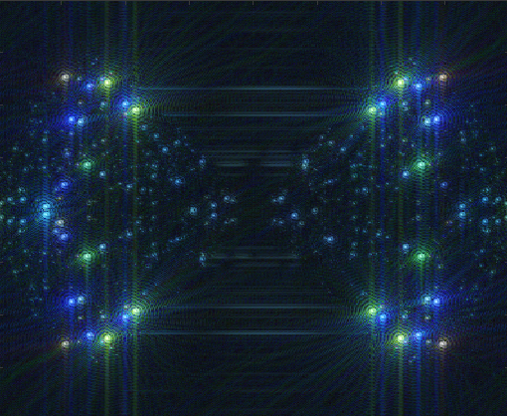} 
        \caption{$a=3; b=15; n=20; m=1.5$.}
        \label{fig12c}
    \end{subfigure}
    \hspace{0.1cm}
    \begin{subfigure}[b]{0.45\textwidth}
        \centering
        \includegraphics[width=0.6\textwidth]{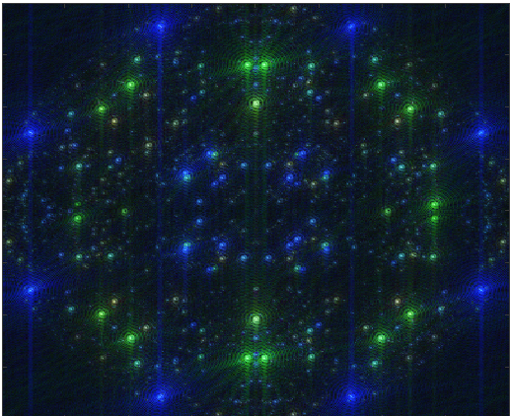} 
        \caption{$a=b=33; n=8; m=1.5$.}
        \label{fig12d}
    \end{subfigure}
     \vfill
    \vspace{0.5cm}
    \begin{subfigure}[b]{0.45\textwidth}
        \centering
        \includegraphics[width=0.6\textwidth]{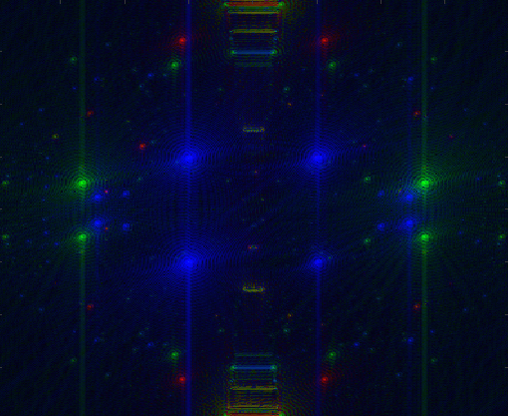} 
        \caption{$a=19; b=86; n=10; m=1.5$.}
        \label{fig12e}
    \end{subfigure}
    \hspace{0.1cm}
    \begin{subfigure}[b]{0.45\textwidth}
        \centering
        \includegraphics[width=0.6\textwidth]{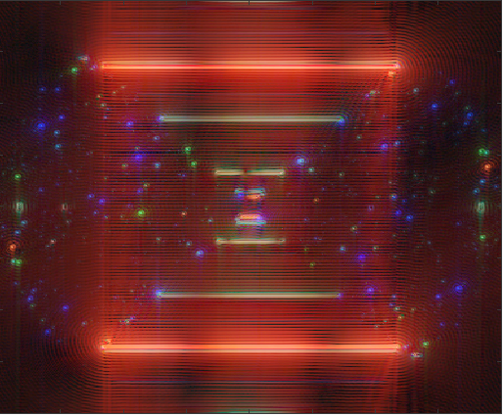} 
        \caption{$a=3; b=18; n=19; m=1.5$.}
        \label{fig12f}
    \end{subfigure}
    \caption{Representative steganograms generated from a second test image using the proposed CHC-based optical encryption and steganographic camouflage algorithm. (a) Original image. (b)--(f) Steganograms obtained with different combinations of the private parameters $(a,b,n)$ while employing CHC-based private keys. The results demonstrate that the proposed method produces a wide variety of visually distinct camouflage patterns while preserving the encrypted information.}
    \label{Fig12}
\end{figure}

Figures~\ref{Fig11} and \ref{Fig12} demonstrate one of the principal characteristics of the proposed methodology. Starting from the same encrypted image, multiple visually distinct steganograms can be generated by modifying either the CHC-based private key or the parameters of the periodic modulation mask. Consequently, the transmitted image no longer exhibits the characteristic appearance of a conventional optical cryptogram, thereby introducing an additional level of uncertainty for an unauthorized observer.


As illustrated in Figs.~\ref{Fig11} and \ref{Fig12}, each camouflaged cryptogram exhibits a distinct visual appearance, even when generated from the same secret image. This variability increases the diversity of the encrypted outputs and makes the detection of encrypted information considerably more difficult.

To recover the original image, an unauthorized user successful recovery of the protected information requires simultaneous knowledge of: (i) obtain the transmitted steganogram, (ii) know the proposed encryption and steganographic camouflage algorithms, (iii) determine the CHC-based private encryption key, and (iv) recover the parameters of the periodic mask $h^{-1}(ax^n,by^n)$. Without all these elements, successful reconstruction of the original image is not possible.

The proposed periodic modulation strategy transforms the cryptogram from a unique encrypted representation into a family of visually distinct steganograms generated from the same protected information. This capability substantially increases the variability of the transmitted images without affecting the authorized reconstruction process, thereby completing the second layer of the proposed optical security framework.

\section{\textbf{Prospective Optical Implementation}}

\begin{figure}
\centering
\includegraphics[width=14cm]{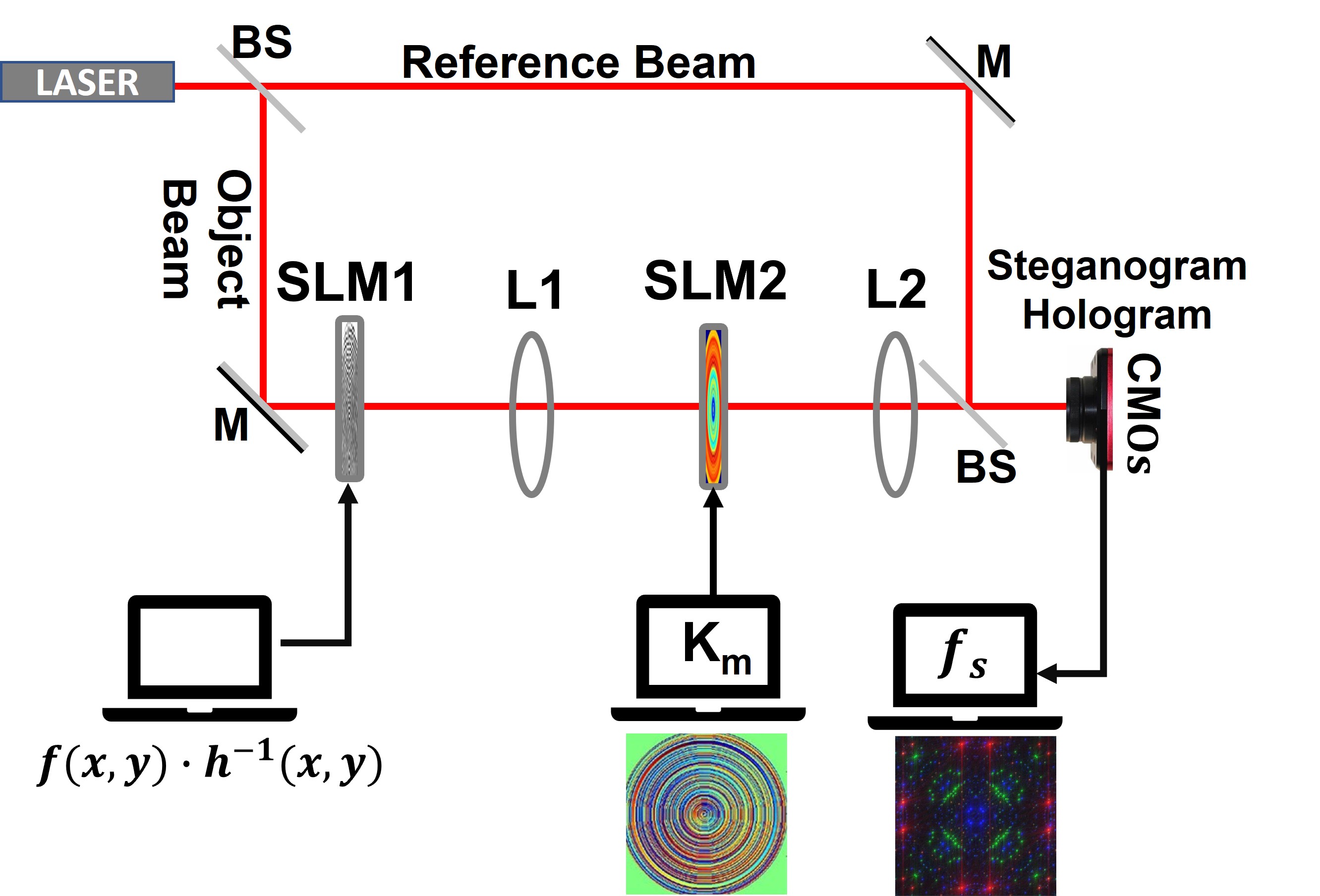} 
\caption{Proposed optical implementation of the CHC-based encryption and steganographic camouflage system using a Mach--Zehnder interferometer. BS: beam splitter; M: mirror; SLM: spatial light modulator; CMOS: digital image sensor.}
\label{fig13}
\end{figure}

The computational formulation presented in the previous sections naturally suggests a practical optical realization. Since all the operations involved in the proposed methodology correspond to well-established optical processing techniques, the complete encryption and steganographic camouflage framework can be implemented using conventional coherent imaging components. Figure~\ref{fig13} illustrates a prospective optical architecture based on a Mach--Zehnder interferometer.

The object arm consists of a conventional $4F$ optical processor. Spatial light modulator SLM$_1$ displays the product
$f(x,y)\,h^{-1}(x,y)$, which is generated digitally prior to optical processing. A second spatial light modulator, SLM$_2$, is positioned at the Fourier plane of lens $L_1$ and displays the CHC-based phase encryption key $K_m(\rho,\phi)$. The optical multiplication performed at the Fourier plane corresponds to Eq.~(\ref{eq:Guv}), while the inverse Fourier transform produced by lens $L_2$ generates the steganographic cryptogram according to Eq.~(\ref{eq:stego}).

In this configuration, the optical processor performs exactly the same sequence of operations described by the computational model, establishing a direct correspondence between the numerical implementation and its prospective experimental realization.

The steganogram is then combined with a coherent reference beam to produce an off-axis digital hologram, which is recorded by a CMOS camera. Digital holographic recording is essential because the steganogram corresponds to a complex optical field whose amplitude and phase must be preserved for faithful reconstruction. Recording only the intensity would prevent accurate reconstruction of the encrypted image during the decryption stage.

To recover the steganogram, the recorded hologram is numerically reconstructed by applying a Fourier transform. The resulting digital hologram constitutes the information transmitted to the authorized receiver.

The same optical arrangement can also be employed for decryption. In this case, the reference beam is blocked, the reconstructed hologram is displayed on SLM$_1$, and the corresponding CHC-based private key is loaded onto SLM$_2$. After optical reconstruction, the recovered field is multiplied digitally by the periodic mask $h(x,y)$ to reveal the original secret image.

Consequently, the proposed architecture provides a feasible optical route for experimentally validating the computational framework presented in this work while preserving the dual-layer security strategy introduced by the proposed methodology.

\section{\textbf{Conclusions}}

This work introduced a dual-layer optical security framework that combines optical encryption based on Circular Harmonic Components (CHCs) with steganographic cryptogram camouflage. Unlike conventional optical encryption systems, in which the cryptogram constitutes the final transmitted object, the proposed methodology transforms the encrypted image into a family of visually distinct steganograms through a second private key based on periodic spatial modulation.

A novel optical encryption framework combining Circular Harmonic Component (CHC)-based keys with steganographic cryptogram camouflage has been presented. The proposed method integrates two complementary security mechanisms: optical encryption protects the information content, whereas steganographic embedding conceals the existence of the encrypted message. Unlike conventional optical encryption schemes that generate white-noise-like cryptograms, the proposed CHC-based representation produces structured cryptograms that can be transformed into a wide variety of visually distinct camouflage patterns by means of a periodic modulation mask.

The proposed algorithm was implemented in MATLAB and evaluated using several test images. The numerical results demonstrated successful encryption and decryption while preserving high reconstruction quality. Furthermore, the proposed approach generated multiple visually different steganograms from the same secret image by varying the private parameters of the periodic mask, thereby increasing the diversity of the transmitted encrypted information without modifying the underlying encryption principle. The proposed optical configuration establishes a direct correspondence between the computational formulation and its future experimental implementation, facilitating the transition from numerical validation to laboratory realization.

An additional contribution of this work is the formulation of a prospective optical implementation based on a Mach--Zehnder interferometer incorporating two spatial light modulators and digital holographic recording. This architecture establishes a feasible pathway toward a hybrid optical--digital realization of the proposed encryption and steganographic camouflage system.

Future work will focus on the experimental implementation of the proposed optical architecture and its validation under real optical conditions. In addition, a comprehensive cryptanalytic assessment of the proposed framework will be conducted, including statistical security metrics, key-space and key-sensitivity analyses, robustness against noise, occlusion, and optical distortions, as well as resistance to known-plaintext and chosen-plaintext attacks. The proposed methodology will also be extended toward real-time optical information processing, enabling practical implementations for secure optical communications and information protection.

Overall, these future developments are expected to consolidate the proposed dual-layer optical security framework as a practical and scalable approach for secure optical communications, optical information processing, and next-generation optical security systems.

\section*{\textbf{Acknowledgements}}
This research was conducted entirely at the Laboratory of Modern Optics Research Group, University of Pamplona. The results are part of the research project "Real-Time Optical Cryptography," funded by the University of Pamplona. Some of the findings are included in the physics undergraduate thesis of the second author and are part of the digital repository of the Virtual Library of the University of Pamplona.

\bibliographystyle{ieeetr}
\bibliography{Bibliografia}

\end{document}